\documentclass[Physsubmission, Phys]{SciPost}

\usepackage{booktabs}
\usepackage{multirow}
\usepackage{tabularx}
\usepackage{float}

\hypersetup{
    colorlinks,
    linkcolor={red!50!black},
    citecolor={blue!50!black},
    urlcolor={blue!80!black}
}

\usepackage[bitstream-charter]{mathdesign}
\DeclareSymbolFont{usualmathcal}{OMS}{cmsy}{m}{n}
\DeclareSymbolFontAlphabet{\mathcal}{usualmathcal}

\begin{document}

\begin{center}{\Large \textbf{
Impact of LHC dijet production in pp and pPb collisions on the nNNPDF2.0 nuclear PDFs
}}\end{center}

\begin{center}
Rabah Abdul Khalek\textsuperscript{1, 2 *}
\end{center}

\begin{center}
\textsuperscript{\bf 1} Department of Physics and Astronomy, \\VU Amsterdam, 1081HV, Amsterdam, The Netherlands
\\
\textsuperscript{\bf 2} Nikhef Theory Group, \\Science Park 105, 1098 XG, Amsterdam, The Netherlands
\\
* rabah.khalek@gmail.com
\end{center}

\begin{center}
\today
\end{center}


\definecolor{palegray}{gray}{0.95}
\begin{center}
\colorbox{palegray}{
  \begin{tabular}{rr}
  \begin{minipage}{0.1\textwidth}
    \includegraphics[width=22mm]{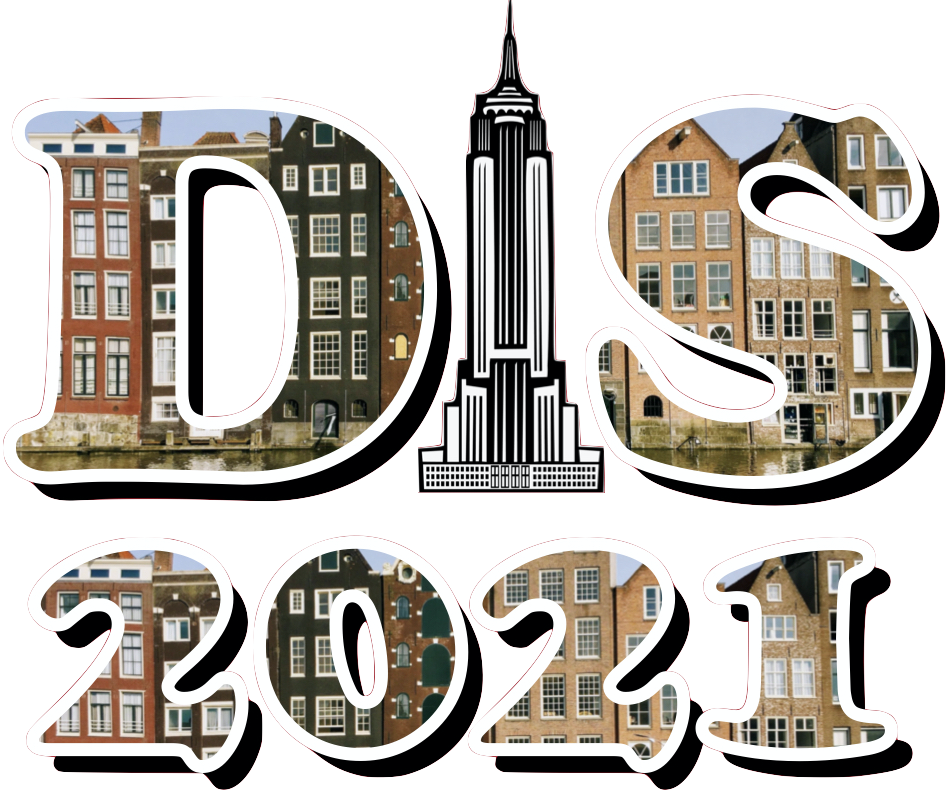}
  \end{minipage}
  &
  \begin{minipage}{0.75\textwidth}
    \begin{center}
    {\it Proceedings for the XXVIII International Workshop\\ on Deep-Inelastic Scattering and
Related Subjects,}\\
    {\it Stony Brook University, New York, USA, 12-16 April 2021} \\
    \doi{10.21468/SciPostPhysProc.?}\\
    \end{center}
  \end{minipage}
\end{tabular}
}
\end{center}

\section*{Abstract}
{\bf
We quantify the impact of LHC dijet production measurements on the nNNPDF2.0 nuclear PDFs in twofold. First, from a proton baseline based on NNPDF3.1 and augmented by pp dijet production measurements from ATLAS and CMS at 7 and 8 TeV. Second, from a new nNNPDF2.0 global analysis including the ratio of pPb to pp dijet spectra from CMS at 5 TeV. We show that as opposed to the CMS at 5 TeV  absolute pp and pPb dijet spectra, the pPb/pp ratio is well described in a nPDFs fit and provides strong constraints on the gluon of lead.
}

\vspace{10pt}

\paragraph{Introduction}
Recent phenomenological and theoretical studies demonstrate that dijet cross section in pp and pPb collisions provides strong constraints on the gluon PDFs~\cite{Giele:1994xd, AbdulKhalek:2020jut} and nuclear PDFs (nPDFs)~\cite{Eskola:2013aya,Paukkunen:2014pha,Armesto:2015lrg,Eskola:2018sxu} respectively. In particular, the relatively small final-state effects (\textit{e.g.} jet quenching) in pPb collisions support the idea of using jets as probes for nPDF fits~\cite{CMS:2014qvs}.

In our previous study~\cite{AbdulKhalek:2020jut}, we investigated the impact of the dijet production measurements from ATLAS and CMS at 7 and 8 TeV~\cite{Aad:2013tea,Chatrchyan:2012bja,Sirunyan:2017skj} on the gluon PDF of NNPDF3.1~\cite{Ball:2017nwa}, compared the results with those of single-inclusive jet, assessed the perturbative behaviour of the observables considered and studied the role played by different factorisation scale choices. In this proceeding, we augment the latter global analysis by the CMS 5 TeV dijet spectra~\cite{CMS:2018jpl} and study the quality of the fit both at NLO and NNLO.

The latest model-independent determination of nPDFs, nNNPDF2.0~\cite{AbdulKhalek:2020yuc}, included various processes that ensured flavour separation. These cover the neutral-current deep-inelastic nuclear structure functions, inclusive and charm-tagged cross-sections from charged-current scattering, all available measurements of W and Z leptonic rapidity distributions in proton-lead collisions from ATLAS and CMS 5 and 8 TeV. In this proceeding, we focus on the ratio of pPb to pp dijet pseudorapidity distributions from CMS at 5 TeV~\cite{CMS:2018jpl} and assess its impact on the gluon of lead.

\paragraph{Experimental data}
The dijet data from CMS at 5 TeV~\cite{CMS:2018jpl} is provided in terms of pseudorapidity distributions of dijets ($\eta_{\text{dijet}} = (\eta_1+\eta_2)/2$) as functions of their average transverse momentum ($p_{T,\text{dijet}}^{\text{avg}}=(p_{T,1}+p_{T,2})/2 \sim Q$) for both pp ($\mathcal{L}_{\text{int}}=27.4\text{ pb}^{-1}$) and pPb ($\mathcal{L}_{\text{int}}=35\text{ nb}^{-1}$) collisions. Three distributions are provided for this data set: the absolute pp and pPb spectra as well as their ratio pPb/pp. The spectra is defined as the number of dijet per bin of $\eta_{\text{dijet}}$ and $p_{T,\text{dijet}}^{\text{avg}}$, normalised by the pseudorapidity-integrated number of dijets in the associated $p_{T,\text{dijet}}^{\text{avg}}$ bin as follows:
\begin{equation}
    {\displaystyle d\left(\frac{1}{N^{\text{col}}_{\text{dijet}}}\frac{dN^{\text{col}}_{\text{dijet}}}{d\eta_{\text{dijet}}}\right)\bigg/dp_{T,\text{dijet}}^{\text{avg}}} \qquad \text{col}=\text{pp, pPb}
\end{equation}
The fiducial cuts~\footnote{We note that the minimal $p_T$ for the leading jet in Ref.~\cite{CMS:2018jpl} is incorrect and the correct cuts are the ones mentioned above.} associated with this measurement are a minimal $p_T$ of $30$ and $20$ GeV for the leading and subleading jets respectively, a distance parameter $R=0.3$ for the anti-$k_T$ recombination algorithm as well as a $2\pi/3$ absolute difference of azimuthal angles between the leading and subleading jets. The only uncertainties associated with this data set are a statistical and systematic uncorrelated uncertainties that are be added in quadrature during the fit.
Based on LO kinematics we can estimate the coverage of this data in terms of the scaling variables $x_{1,2}$ to be:
\begin{equation} \label{eq:dijet_coverage}
    x_{1,2} = \frac{p_{T,\text{dijet}}^{\text{avg}}}{\sqrt{s_{\text{NN}}}}e^{\pm y} \simeq [5\times 10^{-4}, 1]\quad \text{for CMS dijet} \begin{cases} \sqrt{s_{\text{NN}}}&=5020 \text{ GeV}\\ p_{T,\text{dijet}}^{\text{avg}} &\simeq [55,400] \text{ GeV}\\ y &\simeq [-3,3] \end{cases}
\end{equation}

\paragraph{Updates relative to nNNPDF2.0}
In addition to the data sets included in nNNPDF2.0 proton baseline (see Ref.~\cite{AbdulKhalek:2020yuc}), the dijet data sets from ATLAS and CMS at 7 and 8 TeV~\cite{Aad:2013tea,Chatrchyan:2012bja,Sirunyan:2017skj} reviewed in Ref.~\cite{AbdulKhalek:2020jut} are now included. 
In this respect, two new proton baselines\footnote{We will refer to the new baselines with an asterisk, as in nNNPDF2.0$^*$($^1p$).} are determined, one with NLO QCD corrections and the other with NNLO ones implemented by means of K-factors.
The inclusion of the $5$ TeV pp dijet spectra from CMS in a global PDF fit relies on the NLO and NNLO QCD corrections as computed with \textsc{NNLOJET}~\cite{Gehrmann-DeRidder:2019ibf}. Although fast interpolation grids are possible to produce for NLO matrix elements, it is not the case for the NNLO thus the need for K-factors.

\begin{figure}[H]
    \begin{center}
    \includegraphics[width=0.99\textwidth]{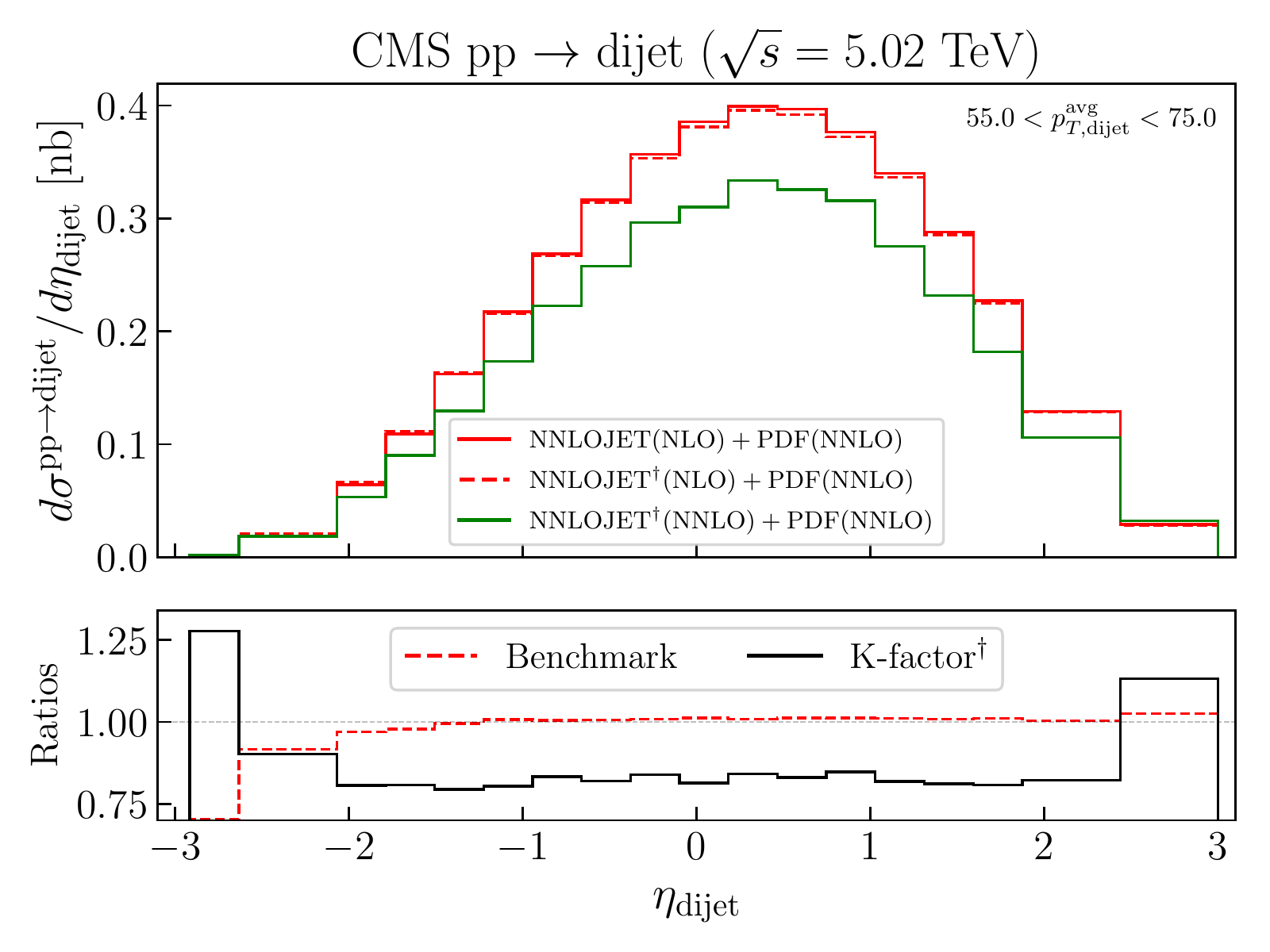}
    \end{center}
    \vspace{-0.8cm}
    \caption{The QCD calculations of the lowest $p_{T,\text{dijet}}^{\text{avg}}$ bin of the $5$ TeV pp dijet spectra from the CMS data set.
        The NLO calculations (solid red histogram) and the reference NLO (dashed red histogram) and NNLO (solid green histogram) calculations using \textsc{NNLOJET} in the upper-panel. The ratio of NLO calculation to the reference ones (dashed red histogram) and NNLO QCD K-factors (solid black histogram). The dagger ($\dagger$) refers to any calculations performed by our co-authors in Ref.~\cite{AbdulKhalek:2020jut}.
    \label{fig:nNNPDF20_CMS_DIJET_kfac}
    }
    \end{figure}

In Fig.~\ref{fig:nNNPDF20_CMS_DIJET_kfac}, we consider a representative bin in $p_{T,\text{dijet}}^{\text{avg}}$ (the lowest among the five available) of the $5$ TeV pp dijet spectra.
In the upper-panel we compare our NLO calculations (solid red histogram) using \textsc{NNLOJET} to the independent calculations at NLO (dashed red histogram) and NNLO (solid green histogram) performed by our co-authors of Ref.~\cite{AbdulKhalek:2020jut} and referred to with a dagger ($\dagger$) in the figure. In the lower-panel we validate the benchmarking (dashed red histogram) where we plot the ratio of our calculations at NLO to the independent calculations. The K-factor (solid black histogram) are also presented. All of these calculations are performed using the NNLO PDF set \texttt{NNPDF31\_nnlo\_as\_0118}~\cite{Ball:2017nwa}.

\begin{figure}[H]
    \begin{center}
    \includegraphics[width=0.99\textwidth]{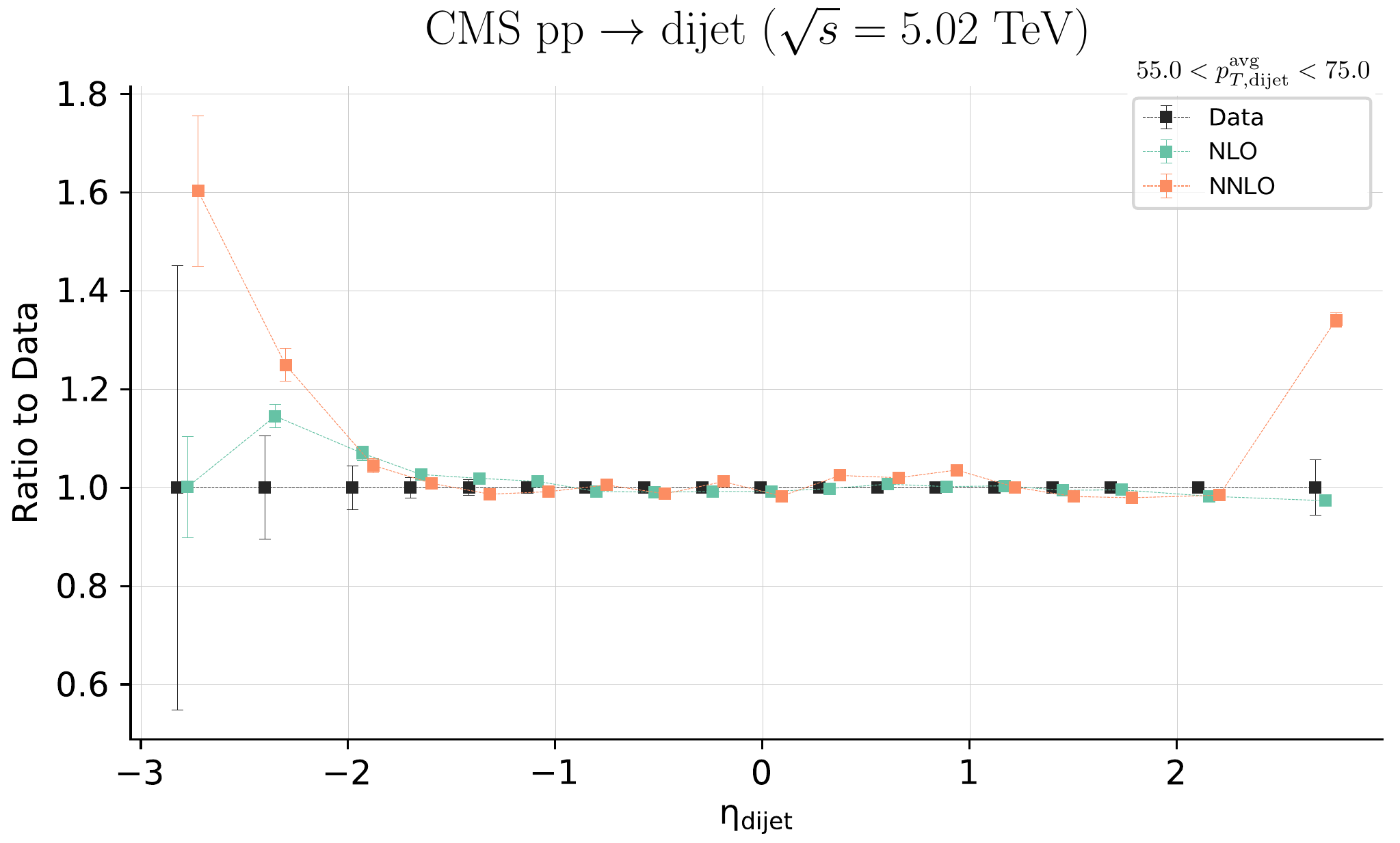}
    \end{center}
    \vspace{-0.8cm}
    \caption{The comparison of  nNNPDF2.0$^*$($^1p$) at NLO (green) and at NNLO (orange) with fitted CMS $5$ TeV pp dijet spectra to the data (black), all normalised to the data points.
    \label{fig:CMS2JET_pp_datavstheory}
    }
    \end{figure}

Table~\ref{tab:nNNPDF20_newbaseline_chi2_1} describes the fit-quality of the new proton baseline nNNPDF2.0$^*$($^1p$) both at NLO and NNLO. We denote by \textbf{w/}(\textbf{w/o}) when the CMS $5$ TeV pp dijet is considered(not considered) on top of the ATLAS and CMS at 7 and 8 TeV data sets.
The proton baseline fits without the CMS $5$ TeV data set follow the same set of conclusions in Ref.~\cite{AbdulKhalek:2020jut} that remain intact with the exclusion of the CHORUS and NuTeV data sets, as well as a lower initial scale ($\mu_0=1$ GeV). At NLO, the description of this data set seems to improve from a $\chi^2$ per data point of $5.87$ to $2.51$ which is between the $\chi^2$ of the CMS at 7 and 8 TeV data sets. However the inclusion of this data set deteriorates the global $\chi^2$, which per data point goes from $1.37$ to $1.42$. At NNLO, the fit quality of this data set also improves upon its inclusion ($12.04$ to $6.91$), however its description is not satisfactory due to the significantly large $\chi^2$. The main contribution to this large $\chi^2$ comes from the extreme pseudorapidity that are harder to fit as shown in the representative Fig.~\ref{fig:CMS2JET_pp_datavstheory}. For this reason, the proton baseline that will be used next is restricted only to the ATLAS and CMS at 7 and 8 TeV data sets.
\begin{table}[H]
    \centering
    \begin{tabular}{lccccc}  \toprule
        & & \multicolumn{2}{c}{NLO} & \multicolumn{2}{c}{NNLO}\\
        data set              & $N_{\text{dat}}$ &  \textbf{w/o}      & \textbf{w/}  &  \textbf{w/o}     & \textbf{w/}\\
        \midrule
        ATLAS 7 TeV             &    90    &   1.03  & 1.01 &   1.98   &    1.91  \\
        CMS   7 TeV             &    54    &   1.58  & 2.03 &   1.75   &    1.92  \\
        CMS   8 TeV             &    122   &   3.87  & 3.61 &   1.48   &    1.55  \\
        \textbf{CMS   5 TeV}    &    \textbf{85}    &  \textbf{[5.87]}  & \textbf{2.51} &  \textbf{[12.04]}    &    \textbf{6.91}  \\
        \midrule
         Total                    &          &  1.37   & 1.42     &   1.24   &   1.41      \\
\bottomrule
\end{tabular}
\caption{The $\chi^2$ per data point for the two new proton baselines: nNNPDF2.0$^*$($^1p$) at NLO and NNLO.
Results are shown
for the dijet data sets together with the number of data points in each
data set.}
\label{tab:nNNPDF20_newbaseline_chi2_1}
\end{table}

Having defined our new proton baselines, we now compare them to the nNNPDF2.0 one. In Table~\ref{tab:nNNPDF20_newbaseline_chi2_2}, we start by comparing the fit quality of the nNNPDF2.0 nPDF sets on the CMS pPb/pp ratio data using the old and new proton baselines (without fitting the data set). The fact that the $\chi^2$ per data point improved (from $3.342$ to $3.145$) merely due to the new proton baseline, highlights on one hand the importance of the proton PDF contribution to the heavy-ion observables and on the other, that the ATLAS and CMS at 7 and 8 TeV dijet data sets in pp provide information that helps describing the pPb/pp CMS $5$ TeV spectra.
\begin{table}[H]
\centering
\renewcommand{\arraystretch}{1.20}
\begin{tabular}{c|ccc}
data set & $N_{\text{dat}}$ &  nNNPDF2.0  & nNNPDF2.0$^*$  \\
& & \multicolumn{2}{c}{NLO} \\
\toprule
CMS dijet pPb/pp 5 TeV & 84 & [3.342]  & [3.145] \\
\bottomrule
\end{tabular}
\vspace{0.3cm}
\caption{\small The $\chi^2$ per data point calculated for the pPb/pp CMS $5$ TeV spectra using the new proton baseline nNNPDF2.0$^*$($^1p$) and the nNNPDF2.0 lead nuclear PDF at NLO. Both values are enclosed in square brackets as the data set is not included in the fit.\label{tab:nNNPDF20_newbaseline_chi2_2}
}
\end{table}

\begin{figure}[H]
\begin{center}
\includegraphics[width=0.99\textwidth]{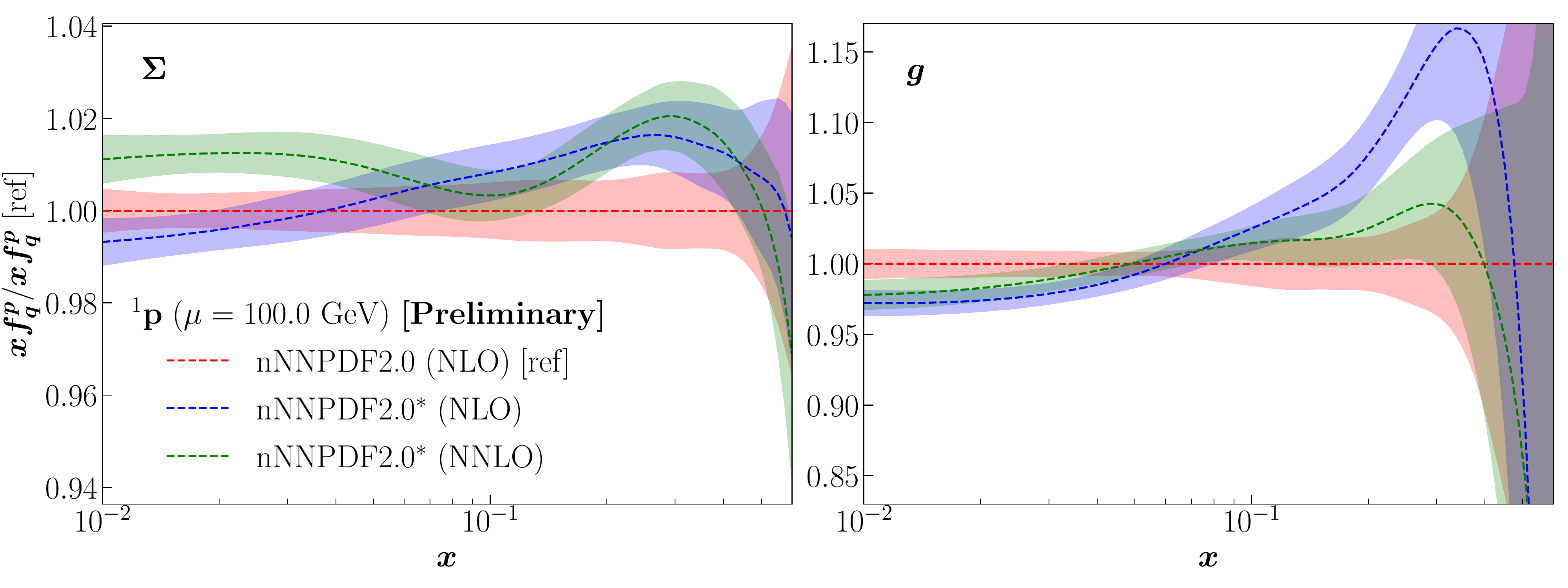}
\end{center}
\vspace{-0.8cm}
\caption{A comparison between the new proton baseline nNNPDF2.0$^*$($^1p$) fitted to the ATLAS and CMS dijet at 7 and 8 TeV and the old nNNPDF2.0 proton baseline.
\label{fig:nNNPDF20_newbaseline}
}
\end{figure}

Finally In Fig.~\ref{fig:nNNPDF20_newbaseline}, we compare the proton PDF baselines themselves. We restrict the flavours to the singlet $\Sigma$ and the gluon and the $x$-range to $[10^{-2},0.6]$ where the NNPDF3.1 set is most sensitive to the new data. As expected, the PDFs show similar trend to those observed in Ref.~\cite{AbdulKhalek:2020jut}, which can be summarised mainly in terms of a reduction of gluon uncertainties at large-$x$.

\paragraph{Results}
We showed that the QCD NNLO calculation did not lead to a satisfactory description of the CMS $5$ TeV absolute pp dijet spectra (see Table~\ref{tab:nNNPDF20_newbaseline_chi2_1}). This turns out to be also the case, even at NLO, for the absolute pPb dijet spectra when augmented to the nNNPDF2.0. For that reason, we focus only on the ratio pPb/pp data that we find to be describable at both NLO and NNLO with a satisfactory $\chi^2$ value. 

In order to gauge the impact of the CMS dijet pPb/pp data set w.r.t. the nNNPDF2.0 data sets in the light of the new proton baseline nNNPDF2.0$^*$($^1p$), we perform a fit at NLO and assess the fit quality in Table~\ref{tab:nNNPDF20_CMS2JET_chi2}. We note that the fit quality of the rest of the data sets (DIS and DY) are very comparable to those quoted in Ref.~\cite{AbdulKhalek:2020yuc}, thus are omitted. This can only mean that this new data set is not in tension with any of the prior data sets considered in nNNPDF2.0. Additionally, Table~\ref{tab:nNNPDF20_CMS2JET_chi2} shows that this new data set is well described when included as a ratio pPb/pp as opposed to the absolute pPb spectra. Upon fitting this data set the $\chi^2$ per data point value goes from a value of $3.145$ to $1.644$ with a global $\chi^2$ per data point of $1.0$. This is mainly due to the cancellation of low-statistics effect and uncertainties from the extreme dijet pseudorapidity bins that we observed in Fig.~\ref{fig:CMS2JET_pp_datavstheory}.
\begin{table}[H]
    \centering
    \renewcommand{\arraystretch}{1.20}
    \begin{tabular}{c|ccc}
        data set & $N_{\text{dat}}$ &  nNNPDF2.0$^*$   & nNNPDF2.0$^*$ + CMS dijet \\
        & & \multicolumn{2}{c}{NLO} \\
        \toprule
        CMS dijet pPb/pp 5 TeV & 84 & [3.145]  & 1.644 \\
        \midrule
        Total & 1551 & [1.192]  & 1.0 \\
    \bottomrule
    \end{tabular}
    \vspace{0.3cm}
    \caption{\small Comparison between the $\chi^2$ per data point of nNNPDF2.0$^*$ and a new NLO determination including the pPb/pp CMS $5$ TeV spectra. Values enclosed in square brackets are of the data set that is not included in the fit.
    \label{tab:nNNPDF20_CMS2JET_chi2}
    }
    \end{table}

\begin{figure}[H]
    \begin{center}
    \includegraphics[width=0.99\textwidth]{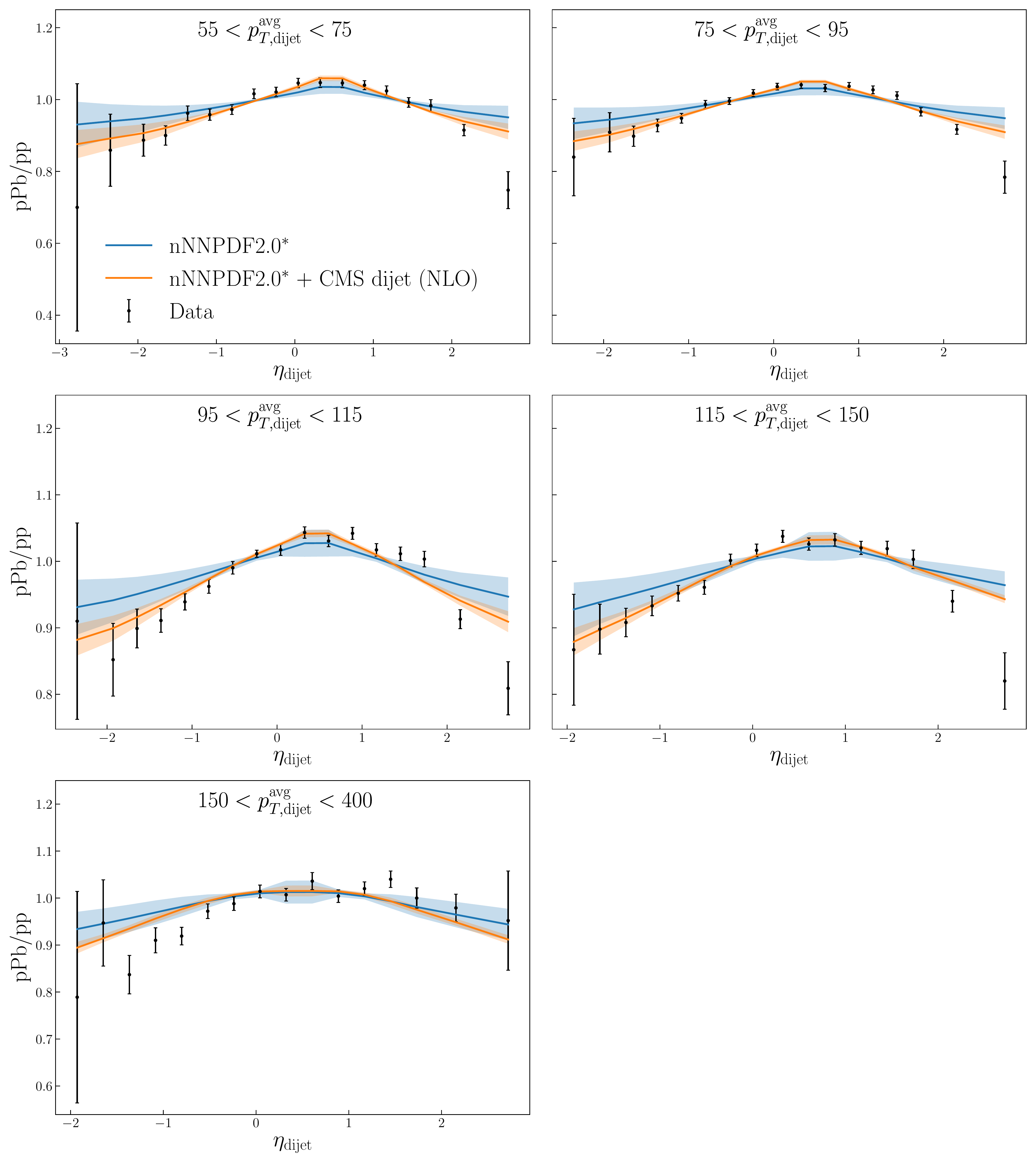}
    \end{center}
    \vspace{-0.8cm}
    \caption{Comparison between the NLO theory predictions of the CMS dijet pPb/pp data from both (nNNPDF2.0$^*$) and (nNNPDF2.0$^*$ + CMS dijet) fits and the data for all bins of $\eta_{\text{dijet}}$ and $p_{T,\text{dijet}}^{\text{avg}}$.}\label{fig:nNNPDF20_CMS2JET_datavstheory}
    \end{figure}

\begin{figure}[H]
    \begin{center}
        \includegraphics[width=0.99\textwidth]{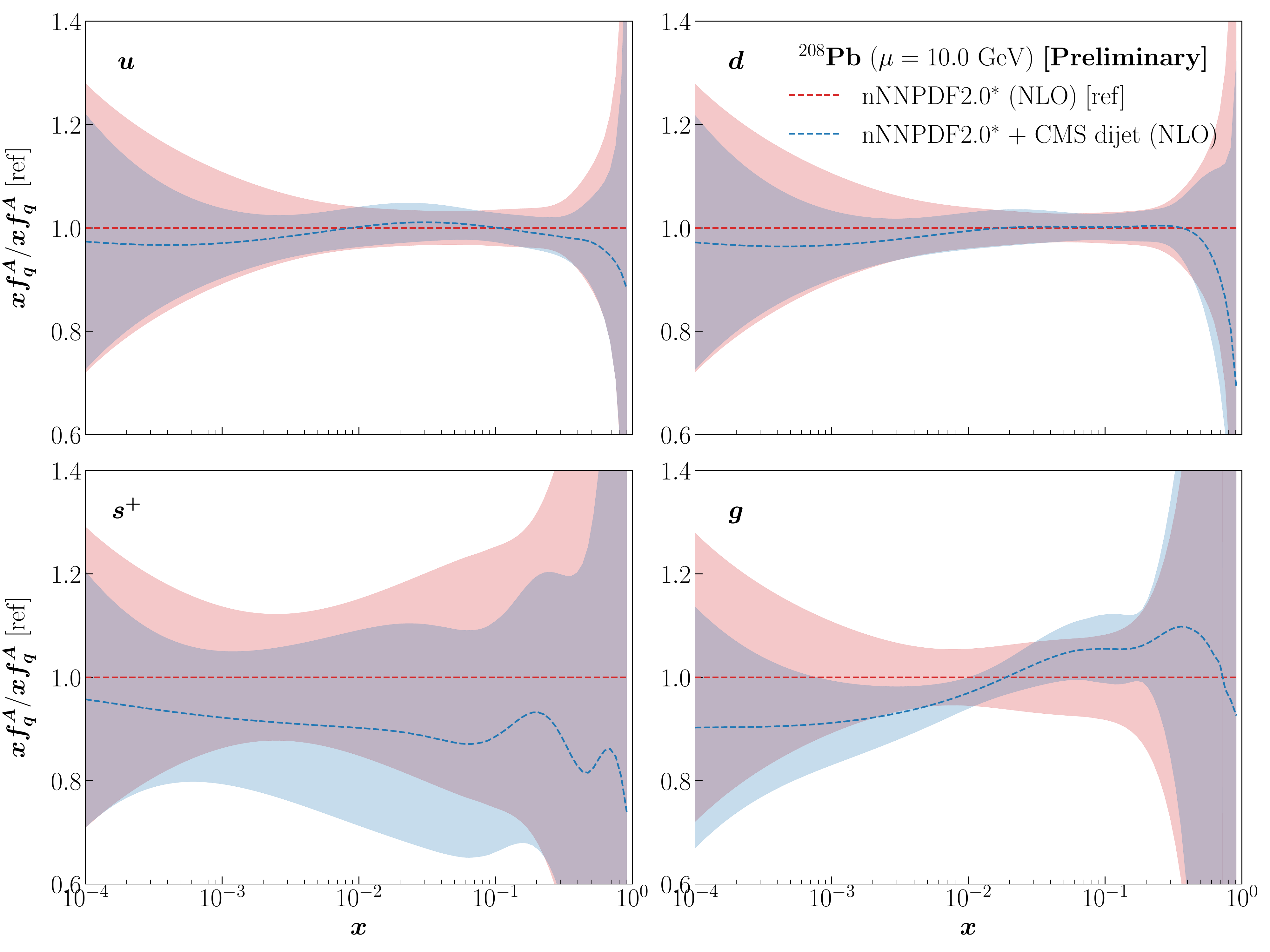}
        \end{center}
    \vspace{-0.8cm}
    \caption{\small Comparison of the up, down, strange and antistrange combination and gluon NLO nPDFs as a ratio of a fit augmented by the CMS dijet pPb/pp data to the nNNPDF2.0-like fit with the new baseline (called nNNPDF2.0$^*$).
        \label{fig:nNNPDF20_CMS2JET}
    }   
    \end{figure}

In Fig.~\ref{fig:nNNPDF20_CMS2JET_datavstheory}, we compare the theory predictions of the CMS dijet pPb/pp data in all bins of $\eta_{\text{dijet}}$ and $p_{T,\text{dijet}}^{\text{avg}}$. One can directly notice that the last $2$ extreme positive $\eta_{\text{dijet}}$ bins in the first $4$ bins of $p_{T,\text{dijet}}^{\text{avg}}$ are the most difficult to fit. Therefore, they must hold the major contribution to the $\chi^2$ per data point of $1.644$. In fact, as we can observe in Table.~\ref{tab:nNNPDF20_CMS2JET_chi2_cut}, upon removing the last $\eta_{\text{dijet}}>2.7$ in all bins of $p_{T,\text{dijet}}^{\text{avg}}$, the $\chi^2$ for the dijet data set reduces from $1.644$ to $1.334$ and the global one from $1.0$ to $0.982$. 
\begin{table}[H]
    \centering
    \renewcommand{\arraystretch}{1.20}
    \begin{tabular}{c|cc|cc}
        data set & $N_{\text{dat}}$ &  CMS dijet & $N_{\text{dat}}$ & CMS dijet ($\eta_{\text{dijet}}<2.7$) \\
        & & NLO & & NLO\\
        \toprule
        CMS dijet pPb/pp 5 TeV & 84  & 1.644 & 79 & 1.334\\
        \midrule
        Total & 1551 & 1.0 & 1546 & 0.982\\
    \bottomrule
    \end{tabular}
    \vspace{0.3cm}
    \caption{\small Same as Table~\ref{tab:nNNPDF20_CMS2JET_chi2}, $\chi^2$ calculated excluding the last $\eta_{\text{dijet}}>2.7$ in all bins of $p_{T,\text{dijet}}^{\text{avg}}$.
    \label{tab:nNNPDF20_CMS2JET_chi2_cut}
    }
    \end{table}
\paragraph{Conclusions} In Fig.~\ref{fig:nNNPDF20_CMS2JET}, we compare the nPDFs obtained from the nNNPDF2.0-like fit with the new baseline (called nNNPDF2.0$^*$) and a different fit augmented by the CMS dijet pPb/pp data. Although nNNPDF2.0$^*$ contains a handful of hadronic data, the new CMS dijet data set provides distinct information, particularly in the $x$-range defined in Eq.~\ref{eq:dijet_coverage}. The most prominent impact is on the gluon, where it's suppressed for $x \leq 10^{-2}$ and enhanced for $ 10^{-2} \leq x \leq 0.3$. 
While not as pronounced as the gluon case, the rest of the plotted flavours also manifest a suppression to accommodate the new data, in particular the combination $s^{+}=s+\bar{s}$. We find these results to be fully compatible with a recent analysis performed by EPPS16 in Ref.~\cite{Eskola:2018sxu}.
%
%

\paragraph{Acknowledgments}
I am grateful for the EPPS16 collaboration, in particular H. Paukkunen and P. Paakinen for providing us with their theoretical predictions for benchmark and for their clarifications regarding the CMS 5 TeV dijet data.
I thank my collaborators J. Rojo and E. R. Nocera for their support with the interpretation of the various results. My work is supported by the Netherlands Organization for Scientific
Research (NWO).


\nolinenumbers

\end{document}